\documentclass[aps,twocolumn,superscriptaddress]{revtex4}
\usepackage{times}
\usepackage[dvips]{graphicx,color}

\newcommand{\ignore}[1]{}

\newcommand{\mComment}[1]{}
\newcommand{\rComment}[1]{}
\newcommand{\gComment}[1]{}
\renewcommand{\mComment}[1]{\textcolor{blue}{Manny: #1}}
\renewcommand{\rComment}[1]{\textcolor{magenta}{Ray: #1}}
\renewcommand{\gComment}[1]{\textcolor{red}{Gerard: #1}}

\newcommand{\ket}[1]{|{#1}\rangle}

\newcommand{\kets}[2]{|{#1}\rangle_{{}_{\!\!{#2}}}}

\newcommand{\slb}[2]{{{#1}^{({#2})}}}

\newcommand{\bop}[1]{{\mathbf{#1}}}
\newcommand{\defeq}{:=}
\newcommand{\mod}{{\;\mbox{mod}}}

\begin{document}

\title{Efficient Linear Optics Quantum Computation}
\author{E. Knill}
\email[]{knill@lanl.gov}
\author{R. Laflamme}
\email[]{laflamme@lanl.gov}
\affiliation{Los Alamos National Laboratory, MS B265, Los Alamos, New Mexico 87545}
\author{G. Milburn}
\email[]{milburn@physics.uq.edu.au}
\affiliation{Center for Quantum Computer Technology, Department of Physics, University of Queensland, St. Lucia, Australia}

\begin{abstract}
We investigate the computational power of passive and active linear
optical elements and photo-detectors. We show that single photon
sources, passive linear optics and photo-detectors are sufficient for
implementing reliable quantum algorithms.  Feedback from the detectors
to the optical elements is required for this implementation. Without
feedback, non-deterministic quantum computation is possible. A
single photon source sufficient for quantum
computation can be built with an active linear optical element
(squeezer) and a photo-detector. The overheads
associated with using only linear optics appear to be sufficiently low
to make quantum computation based on our proposal a viable
alternative.
\end{abstract}

\maketitle

\section{Introduction}

Quantum computers promise to greatly increase the efficiency of
solving problems such as factoring large integers~\cite{shor:qc1995a}
and combinatorial optimization~\cite{grover:qc1997a}. One of the
greatest challenges now is to implement the basic
quantum-computational elements in a physical system and demonstrate
that they can be reliably and scalably controlled.  One of the
earliest proposals~\cite{milburn:qc1988a} for implementing quantum
computation is based on encoding each qubit in two optical modes
together containing exactly one photon. The main problem with this
proposal is that it is extremely difficult to non-linearly couple two
optical modes containing few photons. Here we consider the question of
what can be accomplished in principle using various combinations of
only the simplest optical elements: passive linear optics,
photo-detectors, squeezers (an active linear optical element) and
non-deterministic single photon sources.  We show the surprising
result that quantum computation is possible in principle using the
first three of these. Only weak squeezing is required, and this only
for single photon generation.  Alternatively, any non-deterministic
single photon source can be used. The optical elements must be
controllable based on output from the photo-detectors. An alternative
method relying on phase space based encodings, homodyne detection and
an optical non-linearity in state preparation was independently
presented by Gottesman and Preskill~\cite{gottesman:qc2000a}. This
establishes the principle of linear optics quantum computation
(LOQC). Furthermore, the basic elements are testable in today's
laboratories. Our non-deterministic optical gates may find immediate
application in quantum communication, experiments with entanglement
and optical state preparation.  Spin-offs include near-deterministic
quantum teleportation and a parity measurement for photon based
qubits. Scalable quantum computation using our methods 
requires highly efficient photo-detectors, very low loss
short term photon storage and long state preparation times to achieve
the minimum accuracy required for reliable quantum
computation. However, the theoretical overheads are
sufficiently small compared to the implementation-independent
requirements for scalability to suggest that LOQC is a viable proposal
for implementing quantum computers.

We begin by briefly introducing the basic notions of bosonic modes and
linear optics; give the representation of qubits using bosonic
modes; describe the basic techniques required for implementing
non-deterministic and and then deterministic quantum computation.  We
then consider scalability issues and argue that using quantum 
codes, the complexity of LOQC can be significantly reduced. We conclude
with a discussion of the practical issues involved in using our
methods. We assume some familiarity with quantum optics and quantum
computation.  An introduction to quantum optics can be found
in~\cite{walls:qc1994a}.  An overview of quantum computation and
implementation issues is
in~\cite{aharonov:qc1998a,divincenzo:qc2000a}.

\section{Bosonic modes}

To avoid issues surrounding the polarization states of photons we cast
our work in terms of non-interacting spin-less bosons.  Thus, the
physical system of interest consists of a number of bosonic modes. A
bosonic mode (or quantum harmonic oscillator) is a quantum system
whose state space is spanned by the number states
$\ket{0},\ket{1},\ket{2},\ldots$. The modes are labeled and, with
explicit labels, a number state of mode $l$ is denoted by
$\kets{k}{l}$.  The vacuum state, which satisfies that all modes are
in state $\ket{0}$, is denoted by $\ket{\bop{0}}$.  The observables
for mode $l$ can be constructed from the annihilation operator
$\slb{\bop{a}}{l}$, which satisfies $\slb{\bop{a}}{l}\kets{k}{l} =
\sqrt{k}\kets{k-1}{l}$ for $k\geq 1$ and $\slb{\bop{a}}{l}\kets{0}{l}
= 0$.  $\slb{\bop{a}}{l}^\dagger$ and $\slb{\bop{n}}{l} \defeq
\slb{\bop{a}}{l}^\dagger\slb{\bop{a}}{l}$ are the creation and the
number operator for this mode, respectively.

\section{Optical elements}

The most readily implementable processes are those given by passive
linear optical elements. These are elements whose effect on the state
of the mode are given by the following two Hamiltonians:
\begin{equation}
\begin{array}{rcll}
\slb{\bop{n}}{l}&\defeq&\slb{\bop{a}}{l}^\dagger\slb{\bop{a}}{l}
                &\mbox{(phase shifter)}\\
\slb{\bop{b}}{lm}&\defeq&\slb{\bop{a}}{l}\slb{\bop{a}}{m}^\dagger
                         +\slb{\bop{a}}{l}^\dagger\slb{\bop{a}}{m}
                 &\mbox{(beam splitter)}.
\end{array}
\end{equation}
Evolutions implementable by passive linear optics preserve
the total number of bosons in the modes. It is therefore convenient
to describe them by their effect on the creation operators.
Specifically, if $U$ is the unitary operator associated
with the evolution, then $U$ takes the state
$\slb{\bop{a}}{l}^\dagger\ket{\bop{0}}$ to
\begin{eqnarray}
U\slb{\bop{a}}{l}^\dagger\ket{\bop{0}} &=&
 U\slb{\bop{a}}{l}^\dagger U^\dagger \ket{\bop{0}}\\
   &=&
 \sum_{m}U_{ml}\slb{\bop{a}}{m}^\dagger\ket{\bop{0}},
\end{eqnarray}
using the fact that $U^\dagger\ket{\bop{0}}=\ket{\bop{0}}$.  The
matrix defined by the coefficients $U_{ml}$ must be unitary, and
furthermore, for all unitary $U_{ml}$, there is a sequence of phase
shifter and beam splitter evolutions that implements the
corresponding operation up to a global phase~\cite{reck:qc1994a}. For
a named optical element $X$, let $U(X)$ be the unitary matrix
associated with $X$ according to the above rules.  The unitary
matrices associated with phase shifters $\slb{P_\theta}{l}$ and beam
splitters $\slb{B_\theta}{lm}$ are:
\begin{eqnarray*}
U(\slb{P_\theta}{1}) &=& e^{i\theta}\\
U(\slb{B_\theta}{12}) &=& \left(\begin{array}{ll}
                      \cos(\theta)&-\sin(\theta)\\\sin(\theta)&\cos(\theta)
                      \end{array}\right).
\end{eqnarray*}

General linear optical elements have Hamiltonians that
consist of terms at most quadratic in the annihilation and creation
operators. We will make use of the squeezer, an active
linear element whose Hamiltonian is
\begin{equation}
\slb{\bop{s}}{lm} \defeq \slb{\bop{a}}{l}\slb{\bop{a}}{m}
                         +\slb{\bop{a}}{l}^\dagger\slb{\bop{a}}{m}^\dagger.
\label{eq:squeezer}
\end{equation}
A mode can be prepared in its vacuum state $\ket{0}$.  A
non-deterministic single boson source prepares a given mode in the
state $\ket{1}$ with some probability of success. It is assumed that
the information on whether the source was successful is available
classically.

Measurement is accomplished by means of a particle detector, which
destructively determines whether one or more bosons were present in a
mode. We assume that measurement results can be used to control other
optical elements.  A more powerful detector is a destructive particle
counter, which returns the number of bosons present in a mode.  An
approximate particle counter that suffices for our purposes can be
designed by using $N$ particle detectors operating on $N$ modes
$l_1,\ldots,l_N$. To measure mode $l$, use passive linear optics to
transform 
$
\slb{\bop{a}}{l}^\dagger\rightarrow
   {1\over\sqrt{N}}\sum_{m=1}^{N}\slb{\bop{a}}{l_m}^\dagger,
$
Each of the $N$ modes is then measured using the corresponding
particle detector. If mode $l$ has $k$ bosons, then the probability
that some output mode has two or more bosons is at most
${1-{(N)_k\over N^k}}\leq{k(k-1)\over 2 N}$, where we define $(N)_k$ as the
$k$'th falling factorial of $N$, given by $N(N-1)\ldots (N-k+1)$.
Provided that the maximum number of photons is not too large,
the number of bosons is, with high probability, the number
of particle detectors that detect a boson. The bound
on the error probability is required for estimating
the resources needed to achieve sufficiently high accuracy
for reliable quantum computation.

\section{Representing qubits}

A qubit is a quantum system with two distinguished basis states,
$\kets{0}{q}$ and $\kets{1}{q}$. Thus, the state space of a qubit
generalizes that of a classical bit by allowing superpositions of the
classical states.  To represent a qubit using bosonic modes requires
mapping the basis states unitarily into two states of one or more
modes.  We choose the traditional encoding using two modes and one
boson, where $\kets{0}{q}\rightarrow \kets{0}{a}\kets{1}{b}$ and
$\kets{1}{q}\rightarrow \kets{1}{a}\kets{0}{b}$, and
call this a ``bosonic qubit''. For example, the two
modes can be two orthogonal polarization states of an optical mode, so
that the polarization state of a photon encodes the state of a qubit.
With this encoding, the one qubit ($\mathit{U}(2)$) rotations are
easy to implement using passive linear optics. In the case of encoding
in polarization states, polarization rotators can be used
in place of beam splitters.

To complete the implementation of a standard quantum computer, it is
necessary be able to prepare the state of a qubit in $\kets{0}{q}$, to
measure a qubit in the standard basis $\kets{0}{q},\kets{1}{q}$, and,
in addition to the rotations above, to implement a controlled-not
($\mbox{c-not}$) and/or a controlled-sign ($\mbox{c-sign}$)
gate~\cite{divincenzo:qc2000a}.  The state preparation is accomplished
here by using a single boson source to prepare mode $a$ in
$\kets{1}{a}$ and mode $b$ in $\kets{0}{b}$.  To measure a bosonic
qubit it suffices to use a photo-detector on the first mode.
Unfortunately, linear optics is insufficient for implementing the two
qubit gates $\mbox{c-not}$ and $\mbox{c-sign}$, which are defined in
terms of unitary matrices acting on the four dimensional two-qubit
state space by
\begin{eqnarray}
\mbox{c-not}&=&
\left(\begin{array}{llll}
1&0&0&0\\
0&1&0&0\\
0&0&0&1\\
0&0&1&0
\end{array}\right)
\\
\mbox{c-sign}&=&
\left(\begin{array}{llll}
1&0&0&0\\
0&1&0&0\\
0&0&1&0\\
0&0&0&-1\\
\end{array}\right),
\end{eqnarray}
where the ordering $\ket{00},\ket{01},\ket{10},\ket{11}$ is used for
the basis. Thus $\mbox{c-not}$ flips the second qubit conditional on
the first, and $\mbox{c-sign}$ flips the sign of the second qubit
conditional on the first.  A direct bosonic implementation using a
suitable Hamiltonian requires strong optical non-linearities. For
example the $\mbox{c-sign}$ gate could be implemented with a
non-linear phase shift on the second halves of two bosonic qubits. The
task of this work is to show how these operations can be induced using
only single particle sources and particle detectors.

We conclude this section by introducing one more useful
gate, the Hadamard transform on one qubit, defined by the matrix
\begin{equation}
H = 
{1\over\sqrt{2}}\left(\begin{array}{ll}
1&1\\1&-1
\end{array}\right).
\end{equation}
$H$ can be implemented by a balanced beam splitter on bosonic qubits
(to within a global phase). Note that $\mbox{c-not}$ can be obtained
from $\mbox{c-sign}$ by conjugating the second qubit with the Hadamard
transform (which is its own inverse).

\section{Non-deterministic quantum computation}

A non-deterministic operation is one that succeeds only with
probability $<1$, and whether it succeeded or not is known after it
has been applied.  It turns out that it is much easier to implement a
non-deterministic non-linear phase shift rather than a deterministic
one. These non-deterministic gates will also play a crucial role in
preparing the states required for deterministic non-linear phase
shifts. At the same time, having such a gate implies the ability to
perform non-deterministic quantum computation (conditional on an
acceptable measurement outcome at the end), general state preparation,
and entanglement sharing through lossy systems.  The basic
nondeterministic gate we construct performs the transformation
\begin{eqnarray}
\mathit{NS}_1:\alpha_0\ket{0}+\alpha_1\ket{1}+\alpha_2\ket{2}
 &\rightarrow& \alpha_0\ket{0}+\alpha_1\ket{1}-\alpha_2\ket{2}\nonumber\\
 &&\mbox{with probability 1/4}.
\end{eqnarray}
In order to implement the conditional sign flip $\mbox{c-sign}$
on two bosonic qubits encoded in modes $1,2$ and $3,4$, respectively
do the following: Apply the balanced beam splitter $B_{\pi/2}$
to modes $1$ and $3$. In the case where the two qubits
are both in state $\kets{1}{q}$, modes $1$ and $3$
are in the state $\kets{11}{13}$, which transforms to
${1\over\sqrt{2}}(\kets{20}{13}+\kets{02}{13})$. In none of the
other cases do two bosons appear in the same mode. Now
apply $\mathit{NS}_1$ to mode $1$ and then to mode $3$
and undo the balanced beam splitter. This has the desired effect
with probability $1/16$.

The procedure for applying $\mathit{NS}_1$ to mode $1$ begins with
preparing the initial state $\kets{10}{23}$ in ancilla modes $2$ and $3$.
(The labels are chosen for convenience and are arbitrary.)
Next a sequence of beam splitters is used to implement
the symmetric unitary transformation
\begin{equation}
U = \left(\begin{array}{lll}
1-2^{1/2}&2^{-1/4}&(3/2^{1/2}-2)^{1/2}\\
2^{-1/4}&1/2&1/2-1/2^{1/2}\\
(3/2^{1/2}-2)^{1/2}&1/2-1/2^{1/2}&2^{1/2}-1/2
\end{array}\right).
\end{equation}
Finally, modes $2$ and $3$ are measured, and the outcome
is accepted only when the state is found to be $\kets{10}{23}$, that
is, the state is the same as the initial ancilla state.
To check that this works, observe that due to particle conservation,
the conditional state transformation is of the form
\begin{equation}
\alpha_0\ket{0}+\alpha_1\ket{1}+\alpha_2\ket{2} \rightarrow
\lambda_0\alpha_0\ket{0}+\lambda_1\alpha_1\ket{1}+\lambda_2\alpha_2\ket{2}.
\end{equation}
The $\lambda_k$ are given by
\begin{eqnarray}
\lambda_0 &=& U_{22}\nonumber\\
          &=& 1/2\\
\lambda_1 &=& U_{11}U_{22}+U_{12}U_{21}\nonumber\\
          &=& (1-2^{1/2})/2 + 2^{-1/2}\nonumber\\
          &=& 1/2\\
\lambda_2 &=& U_{11}(U_{11}U_{22}+2U_{12}U_{21})\nonumber\\
          &=& U_{11}(U_{11}U_{22}+U_{12}U_{21}) + U_{11}U_{12}U_{21}\nonumber\\
          &=& (1-2^{1/2})/2 +(1-2^{1/2})2^{-1/2}\nonumber\\
          &=& -1/2
\end{eqnarray}
A direct calculation shows that $U$'s columns are orthonormal.

\section{Near-deterministic quantum computation}

As noted earlier, it is in principle enough to implement a
deterministic $\mbox{c-sign}$ gate. Due to the results from fault
tolerant quantum computing, it is acceptable for the gate to fail with
a small probability (see Sect.~\ref{sect:reliability}).  The basic
idea in implementing a near-deterministic $\mbox{c-sign}$ gate with
low error is to use ideas from quantum
teleportation~\cite{bennett:qc1993b,gottesman:qc1999a}. The trick is
to prepare an appropriate entangled state suitable for teleportation
with the desired gate already applied, before using it for the
teleportation protocol.  The problem then becomes that of preparing
the entangled state (which can be done non-deterministically) and
implementing the requisite measurement in the protocol.

\subsection{Quantum teleportation}

A basic quantum teleportation protocol for transferring the state
$\alpha_0\kets{0}{1}+\alpha_1\kets{1}{1}$ to mode $3$ first adjoins
the ``entangled'' ancilla state
$(\kets{01}{23}+\kets{10}{23})/\sqrt{2}$ to mode $1$. Note that in
this case, the ancilla state is easily generated from $\kets{10}{23}$
by means of a beam splitter.  The second step is to measure modes $1$
and $2$ in the basis
$(\kets{01}{12}\pm\kets{10}{12})/\sqrt{2},(\kets{00}{12}\pm\kets{11}{12})/\sqrt{2}$
(the ``Bell basis'').  We decompose the measurement into two
steps. The first step determines the parity $p$ of the number of
bosons in modes $1$ and $2$ (``parity measurement''). The second
determines the sign $s$ in the superposition. Consider the case where
$p=1$. Then if $s=\mbox{`+'}$, the state of mode $3$ is
$\alpha_0\kets{0}{3}+\alpha_1\kets{1}{3}$. If $s=\mbox{`-'}$, the
state is $\alpha_0\kets{0}{3}-\alpha_1\kets{1}{3}$, which can be
restored to the starting state by using a phase shifter. For $p=0$,
the situation is similar except that $\kets{0}{3}$ and $\kets{1}{3}$
are flipped (and cannot be easily un-flipped using linear optics). The
key property of quantum teleportation is that the input state appears
in mode $3$ up to a simple transformation without having interacted
with mode $3$ after the preparation of the initial ancilla state.

Consider the parity measurement. Applying a balanced beam splitter to
modes $1$ and $2$ and then measuring the number of photons in the two
modes successfully determines the parity, and if it is odd, the
sign. As a result, this measurement method can be used to perform the
teleportation with success probability $1/2$. We refer to this partial
Bell-state measurement as $\mathit{BM_1}$.

Two teleportation steps using $\mathit{BM_1}$ can be used to implement
$\mbox{c-sign}$ with success probability $1/4$. To see how to do this,
observe that to implement $\mbox{c-sign}$ on two bosonic qubits in
modes $1,2$ and $3,4$ respectively, we could try to first teleport the
second modes of each qubit to two new modes (labelled $6$ and $8$) and
then apply $\mbox{c-sign}$ to the new modes.  The full teleportation
procedure requires a correction step that consists of applying a Pauli
operator (nothing, sign-flip, bit-flip, or their product) to each new
mode. Since $\mbox{c-sign}$ is in the normalizer of the group
generated by the Pauli operators, the correction step can be deferred
until after applying $\mbox{c-sign}$. Of course, now there is nothing
preventing us from applying $\mbox{c-sign}$ to the prepared entangled
state before performing the measurement. Thus the implementation of
$\mbox{c-sign}$ is now reduced to the problem of preparing a modified
four mode entangled state $\kets{b'_4}{5678}$, then implementing the
Bell measurement and correcting the resulting state in the new modes.
The modified entangled state is given by
\begin{equation}
\ket{b'_4} = (\ket{1010}+\ket{0110}+\ket{1001}-\ket{0101})/2.
\end{equation}
$\ket{b'_4}$ can be generated with linear optics by following the sequence
\begin{equation}
\begin{array}{rcll}
\ket{01}\ket{01}&\rightarrow&
(\ket{01}+\ket{10})(\ket{01}+\ket{10})/2&\mbox{(with two $B_{\pi/2}$)}\\
&\rightarrow& \ket{b'_4} & \mbox{(with $\mathit{NS}_1$)}.
\end{array}
\end{equation}
The success probability of this procedure is $1/16$, which means that
the expected effort for obtaining one instance of $\ket{b'_4}$
requires $16$ trials of the sequence above.  Once such an instance is
obtained, we can use two $\mathit{BM}_1$ measurements followed by
phase corrections to implement $\mbox{c-sign}$ on two input bosonic
qubits with success probability $1/4$.  To increase the success
probability further we can either change the teleportation procedure or
improve on the parity measurement. We next show how to teleport nearly
deterministically and how to use this for a near-deterministic
$\mbox{c-sign}$. 

\subsection{Near-deterministic quantum teleportation with beam splitters}

An initial entanglement that results in successful teleportation with
probability of success $1/(n+1)$ is given by
\begin{equation}
\ket{t_n} = \sum_{j=0}^{n}\ket{1}^j\ket{0}^{n-j}\ket{0}^j\ket{1}^{n-j},
\end{equation}
where we omitted normalization constants.  The notation $\ket{a}^j$
means $\ket{a}\ket{a}\ldots$, $j$ times.  Note that if we label the
modes $1\ldots 2n$ (left to right), the state exists in the space of
$n$ bosonic qubits, where the $k$'th qubit is encoded in modes $k$ and
$n+k$. Also, $\ket{t_1}$ is the same as the entanglement used in the
basic teleportation protocol of the previous section.

The teleportation protocol using state $\ket{t_n}$ teleports mode $0$
(say) in a superposition of $\ket{0}$ and $\ket{1}$ to one of the last
$n$ modes of $\ket{t_n}$ by a measurement $\mathit{BM}_n$ involving
only the mode to be teleported and the first $n$ modes of $\ket{t_n}$.
The measurement $\mathit{BM}_n$ is implemented using an $n+1$ point
Fourier transform. Let $\hat F_n$ be the unitary matrix
defined by
\begin{equation}
(\hat F_n)_{kl} = \omega^{kl}/\sqrt{n+1},
\end{equation}
where $\omega = e^{i2\pi/(n+1)}$ and $k,l\in 0\ldots n$. $\hat F_n$ is
unitary and therefore implementable with passive linear optics.  Using
the parallel fast Fourier transform (see page 795
of~\cite{cormen:qc1990a}), it can be implemented with $O(n\log(n))$
elements and depth $O(\log(n))$, for $n$ a power of $2$. Alternatively,
a multiport interferometer can be used~\cite{weihs:qc1996a}.  Denote the
optical array for implementing $\hat F_n$ by $F_n$.  To perform
$\mathit{BM}_n$ apply $F_n$ to modes $0\ldots n$ and measure the
number of bosons in each of these modes. Suppose we detect $k$ bosons
altogether. We claim that if $0<k<n+1$, then the teleported state
appears in mode $n+k$ and only needs to be corrected by applying a
phase shift. The modes $2n-l$ are in state $1$ for $0\leq l<(n-k)$ and
can be reused in future preparations requiring single bosons. The
modes are in state $0$ for $n-k<l<n$.  If $k=0$ we learn that the
input state is measured and projected to $\kets{0}{0}$ and if $k=n+1$,
it is projected to $\kets{1}{0}$. The probability of these two events
is $1/(n+1)$, regardless of the input.  We will make use of the fact
that failure is detected and corresponds to measurements in the basis
$\ket{0},\ket{1}$ with the outcome known.  Note that both the
necessary correction and which mode we teleported to are unknown until
after the measurement.

To prove the claim, observe that applying $P_{2\pi k/(n+1)}$ to mode
$k$ for $0\leq k\leq n$ after applying $F_n$ is equivalent to shifting
modes $0\ldots n$ circularly right before applying $F_n$.  This means
that the states $\ket{x_1}=F_n\ket{1}^k\ket{0}^{n-k+1}$ and
$\ket{x_2}=F_n\ket{0}\ket{1}^k\ket{0}^{n-k}$ differ only by phases in
the number basis.  Thus the measurement $\mathit{BM}_n$ cannot
distinguish between the two states.  If the measurement detects $r_j$
bosons in mode $j$ ($\sum_j r_j = k$ in this case), then the relative
phase of the second state is given by $\prod_j \omega^{r_j j}$.
Because of the way $\ket{t_n}$ is entangled with the last $n$ modes,
the measurement outcome transfers a superposition of $\ket{x_1}$ and
$\ket{x_2}$ in the entanglement to a superposition in 
mode $2n-k+1$. The remaining properties of the claim are immediate.

The teleportation trick to improve the probability of success of
$\mbox{c-sign}$ can be used with $\ket{t_n}$ also.  The necessary
modification of $\kets{t_n}{1\ldots 2n}\kets{t_n}{2n+1\ldots 4n}$ to
teleport two modes with an effective application of $\mbox{c-sign}$ is
accomplished by applying $\mbox{c-sign}$ to each pair of modes
$(n+k,3n+l)$ with $k$ and $l$ in $1\ldots n$.  This works up to
qubit sign flips because the output modes that do not receive the
teleported state are in known configurations after the
procedure. Explicitly, the state that needs to be prepared is given by
\begin{eqnarray}
\ket{t'_n} &=&
\sum_{i,j=0}^{n}(-1)^{(n-j)(n-i)}\ket{1}^j\ket{0}^{n-j}\ket{0}^j\ket{1}^{n-j}
\nonumber\\
&&\hspace*{.3in}\times\ket{1}^i\ket{0}^{n-i}\ket{0}^i\ket{1}^{n-i}. 
\end{eqnarray}

We note a few features of this method of implementing $\mbox{c-sign}$
that will prove useful when using quantum error-correction to boost
reliability. The first step in the implementation of $\mbox{c-sign}$
using $\ket{t'_n}$ is to perform $\mathit{BM}_n$ on the first qubit
and the first n modes of $\ket{t'_n}$. With probability $1/(n+1)$, the
measurement fails, and the first qubit is measured in the standard
basis with a known outcome. If this happen we do not attempt to
complete the protocol so that the second qubit remains coherent. If
the first measurement succeeds, the correction step is applied. Note
that as it only adjusts the phase, it commutes with the implemented
operation.  Next $\mathit{BM}_n$ is applied to the second qubit and
the third $n$ modes of $\ket{t'_n}$.  With probability $1/(n+1)$, this
fails and the second qubit is measured in the standard basis with
known outcome. The outcome affects whether or not the first qubit
experienced a sign flip, which can be corrected with a phase shifter
to restore it to its original coherent state.

\subsection{State preparation algorithm}

We have reduced the problem of implementing a near-deterministic
$\mbox{c-sign}$ to generating the state $\ket{t'_n}$.  Clearly this
can be done by first generating two copies of $\ket{t_n}$ and then
applying the $O(n^2)$ $\mbox{c-sign}$ operations.  Since $\ket{t_n}$
is a uniform superposition of simple bosonic qubit states (the unary
numbers from $0$ to $n$), it is not hard to see how to construct a
quantum network with $O(n)$ gates from an initial state accessible
with single boson sources.  Together this gives $O(n^2)$ gates, which
can be implemented non-deterministically. Although $n$ needs to be no
larger than some constant (pessimistically no more than 25, see
Sect.~\ref{sect:reliability}, implying existence of techniques with
asymptotically efficient resource usage), both the number of gates and
the way in which non-deterministic gates are used needs to be
improved. The remainder of this section is dedicated to this task.

We begin by generating a variant of $\ket{t_n}$ with an ancilla
bosonic qubit that contains parity information about the states
in the superposition:
\begin{eqnarray}
\ket{tp_n} &=&  
  \sum_{j=0}^{n}\ket{1}^j\ket{0}^{n-j}\ket{0}^j\ket{1}^{n-j}\nonumber\\
&&\times
            \kets{(n-j)\mod(2)}{a_1}\kets{(n-j+1)\mod(2)}{a_2}.\nonumber\\
\end{eqnarray}
To prepare $\ket{t'_n}$ one can first prepare two copies of
$\ket{tp_n}$,
apply $\mbox{c-sign}$ to the two ancilla qubits, apply
$B_{\pi/2}$ to each ancilla qubit and measure them. There are
four equiprobable outcomes to the measurement, which differ
from the desired state $\ket{t'_n}$ only by signs, which
can be undone by applying $P_{\pi}$ to the last $n$ modes
of one or both halves of the component states. This involves $O(n)$
individual steps, only one of which requires a non-deterministic
element.

To prepare state $\ket{tp_n}$, begin with the state
$(\ket{1}^n\ket{0}^{n}+\sqrt{n}\ket{1}^{n-1}\ket{0}\ket{0}^{n-1}\ket{1})
 (\ket{01}+\ket{10})$,
which can be generated from a product of single boson states by
applying beam splitters to modes $n$ and $2n$ and to the ancilla
modes. Apply $\mbox{c-sign}$ to modes $2n$ and $2n+1$.
Let $q_l$ be the qubit encoded in modes $l$ and $n+l$ and $a$
the ancilla qubit.  The $l$'th step ($0\leq l\leq n-1$) consists of
\begin{itemize}
\item[1] Conditionally on qubit $q_{n-l}$ being in the state $\ket{01}$,
apply ${B_{\theta_l}}$ to qubit $q_{n-l-1}$, with
$\tan(\theta_l) = \sqrt{n-l-1}$. 
\item[2] Apply $\mbox{c-sign}$ to modes $2n-l-1$ and $2n+1$.
\end{itemize}
At the end undo the beam splitter operation on the ancilla qubit.
$O(n)$ operations are required for implementing this algorithm and
$O(n)$ of these are based on gates that we can only do
non-deterministically.  The steps of the algorithm can be implemented
with $\mbox{c-sign}$, beam splitters and phase shifters using (for
example) the methods of~\cite{barenco:qc1995a}.  For example, the
conditional beam splitter is obtained by conjugating a beam splitter on
the target by $\mbox{c-sign}$ and following that with the inverse beam
splitter.  As this requires two $\mbox{c-sign}$
operations, it may be more efficient to directly implement a
conditional phase shift for phases other than $-1$ and then to
conjugate this by a one qubit rotation to implement the required
conditional beam splitter. If this is done by using an instance of our
teleportation schemes (e.g. with $n=1$), then the built in error detection
may be exploited to achieve some error recovery. In particular, of the
two failure modes for a gate implemented by teleportation, one results
in a state that is essentially a $\ket{tp_l}$ for $l<k$.  This state
can be used for improvements in the probability of success for the
operations required when re-attempting the construction of
$\ket{tp_n}$.

The non-deterministic aspects of the method for preparing $\ket{tp_n}$
complicate the resource analysis. Let $S(n)$ be the expected number of
elementary operations needed to prepare $\ket{tp_n}$.  If we do not
attempt to recover from failure, nor exploit improvements in the
probability of success by using $\ket{tp_k}$ for $k<n$, then the
probability of success would be ${1\over 4}^{O(n)}$, leading to $S(n)
= 4^{O(n)}$ exponentially large.

Using $\ket{tp_k}$ recursively leads to a subexponential method as the
following argument shows: Except for differences in rotation angles,
the algorithm for preparing $\ket{tp_n}$ looks like the algorithm for
preparing $\ket{tp_{n-1}}$ followed by $<C$ non-deterministic
operations (for some constant $C$). Suppose we use $2C$
$\ket{tp_{\sqrt{n}}}$ for these operations. Each of these states
requires $S(\sqrt{n})$ operations and their application has an
additional overhead of $\leq D\sqrt{n}$. The success probability is
is now $1-1/\sqrt{n}$. Therefore, $S(n)\leq
(1-1/\sqrt{n})^{-2C}(S(n-1)+2 C (S(\sqrt{n}) +D\sqrt{n}))$.  Under the
assumption that $S(n) = \Omega(\sqrt{n})$, $S(n)\leq
(1+C_1/\sqrt{n})(S(n-1) + C_2 S(\sqrt{n}))$ for suitable choices of
the constants. This implies that $S(n) = 2^{O(\sqrt{n})}$.

\subsection{Near-deterministic loss detection and  non-destructive parity measurement}

Our constructions so far assume essentially perfect optical gates and
detectors. Unfortunately, loss of bosons is one of the primary error
mechanisms. Since this implies ``leakage'' errors for the bosonic
qubits, scalability requires a non-destructive method for detecting
when the modes supporting a qubit are no longer in a state associated
with the bosonic qubit and returning them to a qubit state.
Fortunately, our methods can be adapted to yield a non-destructive
parity measurement for the space spanned by the states
$\ket{00},\ket{10},\ket{01},\ket{11}$. In order to implement such a
measurement, it is sufficient to follow the procedure used for
teleporting the $\mbox{c-sign}$ gate, but using instead of
$\ket{tp_{n}}$ the state
\begin{eqnarray}
\ket{p'_n} &=& \sum_{i,j:
i+j=0(2)}\ket{1}^j\ket{0}^{n-j}\ket{0}^j\ket{1}^{n-j}\nonumber\\
&&\hspace*{.1in}\times
            \ket{1}^i\ket{0}^{n-i}\ket{0}^i\ket{1}^{n-i}.\\
\end{eqnarray}
The measurement determines whether the number of bosons in the two
input modes is even or odd, and the conditional state can be extracted
from the output modes after suitable phase shifts. The state can be
prepared using a simple variation of the method for preparing
$\ket{t'_n}$, using only one ancilla qubit for keeping track of the
total parity. Measurement of the ancilla at the end yields either
$\ket{p'_n}$ or the odd variant, which is equally useful.

The above loss detection method is incomplete for detecting leakage in
one respect: it does not detect when a mode has two or more
bosons. However, it is sufficient for the experimental procedures to
guarantee return to the coding state with sufficiently high
probability, that is, the loss need not be detected by the
user. Although the information is lost in the process without that
loss being detected, this is in principle sufficient for meeting the
scalability requirements. The teleportation methods described earlier
for implementing various gates ensure that each output mode has at
most one boson. By using them sufficiently frequently on both halves
of a bosonic qubit, return to the state space where loss can be
detected is thus adequately assured.

Another application of the non-destructive parity measurement is to
the teleportation of a bosonic qubit using the traditional entangled
state $\ket{E} = \ket{0110}-\ket{1001}$.  The initial state is
$(\alpha_0\ket{01}+\alpha_1\ket{10})\ket{E}$ on modes $1,2,3,4,5$ and
$6$, say. The Bell measurement requires measuring in the basis
$\ket{0110}\pm\ket{1001}$ and $\ket{0101}\pm\ket{1010}$ This has been
implemented experimentally with some probability of
failure~\cite{bouwmeester:qc1998a,boschi:qc1998a,kimble:qc1998a}.
A non-destructive parity measurement on modes $2$ and $3$
determines which of the two pairs of basis states are present.
The sign can then be determined by applying two
balanced beam splitters to modes $1,2$ and $3,4$
before measuring the four modes.

To conclude this section, observe that the entanglement needed for
teleportation can be produced locally, non-deterministically by
performing the non-destructive parity measurement on modes $1,2$ of
the non-traditional entangled state
$(\kets{01}{1,3}-\kets{10}{1,3})(\kets{01}{2,4}+\kets{10}{2,4}$ and
accepting if the parity is odd. Of course, an appropriate entangled
state is still used locally inside the parity measurement, but this
one may be generated locally, perhaps using the techniques of this
paper. The curious feature is that the distribution of the
entanglement requires only two independent bosons sent by means of
beam splitters. The rest is taken care of by classical communication.

\section{Single bosons using active linear optics}

The need for non-linearities to create single boson states can be
eliminated using weak squeezing and particle detectors~\cite{hong:qc1986a}.  A
non-degenerate squeezer on modes $1$ and $2$ applies the Hamiltonian
$\slb{\bop{s}}{12}$ given in Eq.~\ref{eq:squeezer}.  When applied to
the vacuum, the output consists of a superposition of states with identical
numbers of bosons in the two modes.  Under weak squeezing
conditions, the relative amplitudes of states with a total of
$2n$ bosons decrease exponentially with $n$. Thus a single boson can be
produced in mode $1$ conditional on detection of at least one in mode
$2$. The output state has arbitrarily large overlap with the
desired single boson state. The overlap is one in the limit
of weak squeezing. If we can count bosons in mode $2$, the
reliability of the conditional output is even better.

\section{Reliable scalability}
\label{sect:reliability}

That quantum computing is scalable in principle is a consequence of
the accuracy threshold
theorem~\cite{aharonov:qc1996a,kitaev:qc1997a,knill:qc1998a,preskill:qc1998a}.
According to this theorem, the main requirement for efficient
scalability of a physical implementation of quantum computation is
that the elementary operations can be implemented with a minimum
accuracy, which is currently estimated to be less than
$1-10^{-4}$~\cite{gottesman:qc2000b}. Although achieving such accuracy
may seem like a daunting task, in practice, errors behave much more
predictably than assumed in the general analyses, and experience shows
that at least some aspects of many quantum experiments are
controllable with accuracies above the threshold. A typical example is
the phase of RF pulses in nuclear magnetic resonance. Such accuracies
can be exploited to ``boost'' the accuracy of other gate parameters.

Achieving the necessary accuracy for the methods introduced in this
work is clearly not practical without additional work.  For example,
to implement a single two qubit gate with error rate $<10^{-4}$ would
require $2*10^4$ modes, half of which are initially loaded with single
bosons. To get the necessary prepared state would require a very large
number of tries indeed. To avoid doing this we take advantage of the
fact that ideally, it is always known when a gate fails. Thus, we
begin by assuming that all optical operations (beam splitters, phase
shifters, single boson state preparation, number state measurement)
are perfect. Under this assumption, the threshold can be greatly
improved by exploiting quantum erasure codes~\cite{grassl:1996a},
yielding an ``erasure threshold''. Given the special nature of the
detected errors in our system, we are actually interested in an even
more benign threshold $T_d$. This results in quantum code enhanced
systems, where qubits are highly protected in specially coded
states. One can then estimate how errors in the operations propagate
to errors in specific instances of such systems and apply the general
threshold to obtain an estimate of the minimum accuracy required.
Most importantly, the overhead associated with using linear optics is
directly related to $T_d$ via the complexity of the necessary erasure
code implementation.  We pessimistically estimate $T_d$ to be well
below $.96$, which implies that scalability is achievable using our
teleportation protocols with size parameter $n\leq 25$. This still
implies substantial overheads, albeit far from what might have been
expected.  The quantum code constructions that lead to our estimate
of $T_d$ and more detailed resource analyses are
in~\cite{knill:qc2000a}.

\section{Experimental requirements}

A crucial question is to what extent the methods of the previous
sections can be implemented experimentally. Although large scale
quantum computation is clearly still out of reach with current
technology, many of the elements can be tested using existing
equipment.  For example, the non-deterministic $\mbox{c-sign}$
requires only three modes and one ancilla single photon. Similarly,
the simplest methods for generating highly entangled states or for
teleportation with parity measurements require reasonable overheads.

It is relatively easy to couple two optical modes at a single beam
splitter. As subsequent beam splitters are added however careful
spatial and temporal mode matching is required, otherwise unwanted
modes could be mixed in representing loss channels.  Fortunately a
four mode experiment could be achieved using only two spatial modes
together with the two polarization degrees of freedom for each. The
price would be to add a requirement for polarizing beam splitters,
together with controllable polarizers (Faraday rotators), to enable
flexibility in linear coupling and to separate the polarization modes
for detection.

The requirement of single photon sources is a more difficult
constraint to satisfy with available technology. To some extent this
is mitigated by the conditional preparation of the non deterministic
phase shift; while the time at which the gate is applied is random, we
do know when it has occurred.  One way around this difficulty is to
use a pulse-trigger to indicate that temporally correlated photon
pairs are incident to the device. For example, in the case of the
non-deterministic $\mbox{c-sign}$ method, a three mode state with a
total of three photons, together with an ancilla mode in the vacuum
state is required.  The three mode state can be generated using a
method similar to that used to generate a GHZ
state\cite{bouwmeester:qc1999a} via a type II parametric down
converter with a suitable arrangement of polarizing beam splitters. To
second order in the pump amplitude, the result is a four mode state
with at most four photons in total. Conditioning on a single count in
a time window for one of the four outputs distributes three photons
across the other three modes. Subsequent processing could then yield
the required input states to the non deterministic c-sign
protocol. The considerable disadvantage of this approach is the very
low probability of required states per run.  Single photon sources
however are not far away. The turnstile proposal of~\cite{kim:qc1999a}
or the SAW method of~\cite{foden:qc2000a} could yield pulses of light
with a predefined maximum number of photons per pulse, synchronized to
an electronic clock signal. Optical delay line methods could then be
used to bring single photon pulses together at a particular time and
place. Of course single photon detectors with high quantum efficiency
will be required in all cases.

In an experiment it would be necessary to be able to distinguish that the
required controlled phase shift had been implemented in a successful
run. The simplest way to do this is to use a double path
interferometer with a controlled phase shift introduced in one arm
only. The input signal state is then split at a 50/50 beam splitter at
the input to the device, passes along two arms and is then recombined
at an output beam splitter.  In the absence of a phase shift the
interferometer could be adjusted to transmit the input state with unit
probability at a single output port.  When the phase shift is
introduced however this probability would change as ``which path''
information is imprinted by the conditional phase shift. In essence
the total experiment, including the conditional state preparation, is
a kind of four photon coincidence experiment.

\section{Discussion}

The ability to implement quantum computation with linear optics and
particle detectors realizes the dream of computing with
non-interacting particles.  The only particle interactions occur
implicitly in the detectors and result in particle destruction.  The
present work shows how to do this for bosons, but similar techniques
work for fermions, which have the property that the creation operators
satisfy anti-commutation laws. This greatly improves on previous
methods for implementing quantum networks with linear
optics~\cite{cerf:qc1998b,howell:qc2000a,kwiat:qc2000a}, which require
an exponential number of modes to represent the state space of $n$
qubits. It solves a problem we first learned from Paul Kwiat, who
asked what the computational power of prepared entanglement in optics
is.  Other studies of quantum computation with harmonic oscillators or
continuous variables have depended on non-linear effects either for
quantum
gates~\cite{milburn:qc1988a,howell:qc2000b,braunstein:qc1998a,lloyd:qc1998b}
or for state preparation~\cite{gottesman:qc2000a}.

LOQC is related to the observation that in the usual model of quantum
computation, it suffices to be able to implement operations in the
normalizer of the Pauli group, provided one non-stabilizer state can
be generated, such as the state
$\cos(\pi/8)\ket{0}+\sin(\pi/8)\ket{1}$~\cite{knill:qc1998a}, or a
non-stabilizer measurement can be performed, such as that of the
Hadamard operator $H$.
Linear optical elements generate the normalizer of the continuous
version of the Pauli group. This group is given by products of
operators of the form $e^{-it(\bop{a}+\bop{a}^\dagger)}$ and
$e^{-t(\bop{a}-\bop{a}^\dagger)}$, which correspond to translations in
phase space. From this point of view, the vacuum state preparation
and particle detectors serve the
purpose of the non-stabilizer state and measurement, respectively.

Even though LOQC is no more practical given current technology than
any of the other proposals, many of the basic elements can be tested,
and there are numerous other applications in quantum information
processing.  For example, our methods can be used to generate complex
entangled states with higher efficiency than is possible by using
sequential down-conversion sources. The non-destructive parity
measurement can be used for implementing standard quantum
teleportation essentially unconditionally. In general our methods
yield complex quantum non-demolition measurements.  Since quantum
communication is substantially less demanding than quantum computation
(as demonstrated by its much better
thresholds~\cite{briegel:qc1998a}), our methods are closer to being
practical for implementations of quantum channels, particularly when
relatively low success probabilities are acceptable, such as in
quantum cryptography. Note that quantum optics is likely to dominate
the important communication sector of quantum information processing
regardless of how quantum computation is finally implemented.

Another approach to implementing quantum computation optically is to
use non-linear optical elements. Assuming that such elements are
available, though perhaps with high losses, it may be possible to
combine techniques from LOQC with these
elements to significantly reduce the introduced errors.

More work is required to optimize the resources required for LOQC.
Clearly the suggestions made here are just the beginning and we
believe that significant improvements are possible. From a theoretical
point of view it would be nice to establish lower bounds on the
complexity of various elementary gates depending on the desired
probability of failure.

Acknowledgments: We thank the Aspen Center for Physics for its
hospitality. We are grateful to Paul Kwiat for his suggesting the
problem to us and for his help with the references.  E.K. and
R.L. received support from the NSA and from the DOE (contract
W-7405-ENG-36).

\end{document}